\documentclass[letterpaper,11pt]{article}
\pdfoutput=1

\usepackage{jheppub}
\usepackage{multirow}

\usepackage{subfig}
\usepackage{xspace}
\usepackage[countmax]{subfloat}

\usepackage{amssymb}
\usepackage{amsmath}
\usepackage{cancel}
\usepackage{color}
\usepackage{braket}
\usepackage{graphicx}
\usepackage{multirow}
\usepackage{verbatim}
\usepackage{amsthm}
\usepackage{slashed}
\usepackage{wasysym}
\usepackage{simplewick}
\usepackage{mathtools}
\usepackage{soul}
\usepackage{xspace}

\newcommand{\Tau}{\mathcal{T}}

\newcount\colveccount
\newcommand*\colvec[1]{
        \global\colveccount#1
        \begin{pmatrix}
        \colvecnext
}
\def\colvecnext#1{
        #1
        \global\advance\colveccount-1
        \ifnum\colveccount>0
                \\
                \expandafter\colvecnext
        \else
                \end{pmatrix}
        \fi
}

\newcommand{\refcite}[1]{ref.~\cite{#1}}
\newcommand{\refscite}[1]{refs.~\cite{#1}}
\newcommand{\eq}[1]{eq.~\eqref{eq:#1}}

\newcommand{\secs}[2]{secs.~\ref{sec:#1} and \ref{sec:#2}}

\newcommand{\fig}[1]{fig.~\ref{fig:#1}}
\newcommand{\figs}[2]{figs.~\ref{fig:#1} and \ref{fig:#2}}

\newcommand{\Msquared}{A}                  
\newcommand{\as}{\alpha_s}

\def\cM{\mathcal{M}}

\def\cO{\mathcal{O}}

\def\nn{{\nonumber}}

\DeclareRobustCommand{\Sec}[1]{Sec.~\ref{sec:#1}}

\def\be{\begin{equation}}
\def\ee{\end{equation}}

\def\LO{\text{LO}}

\newcommand{\tauc}{{\tau_{\text{cut}}}}
\newcommand{\taucat}[1]{{\tau^{#1}_{\text{cut}}}}
\newcommand{\dsigmadtau}{\frac{\df \sigma}{\df \tau}} 
\newcommand{\dsigmadtauat}[1]{\frac{\df \sigma^{#1}}{\df \tau}} 

\allowdisplaybreaks[3]

\newcommand{\df}{\mathrm{d}}

\newcommand\bn{{\bar n}}

\newcommand{\eps}{\epsilon}

\newcommand{\lep}{\mathrm{lep}}
\newcommand{\hadcm}{\mathrm{cm}}

\newcommand{\Obs}{\cO}
\newcommand{\Obst}{\tilde\cO}

\newcommand{\Ecm}{E_\mathrm{cm}}

\newcommand{\NLP}{{\rm NLP}}

  \newcount\hour \newcount\minute
  \hour=\time \divide \hour by 60 \minute=\time
  \newcommand{\todaytime}{\today \ -- \number\hour :\ifnum \minute<10 0\fi\number\minute}

\preprint{CERN-TH-2024-002}

\title{N$^3$LO Power Corrections for $0$-jettiness Subtractions With Fiducial Cuts}

\abstract{We compute the leading logarithmic power corrections at next-to-next-to-next-to-leading order for $0$-jettiness subtractions for color singlet production. 
We discuss how to disentangle these power corrections from those arising from the presence of fiducial and isolation cuts by using Projection-to-Born improved slicing.
We present the results for Drell-Yan and Higgs production in gluon fusion differential in both the invariant mass and rapidity of the color singlet.
Our results include all the channels contributing at leading logarithmic order for these processes, including the off-diagonal channels that receive contributions from soft quark emission. 
We study the numerical impact of the power corrections for Drell-Yan and Higgs production and find it to become negligible compared to the size of the N$^3$LO corrections only below $\tau_\text{cut} \sim 10^{-5}$. 
We estimate that in a fully differential calculation at N$^3$LO combining the Projection-to-Born improved slicing method and our results for the leading logarithmic power corrections may allow for keeping the slicing uncertainties under control already with $\tau_\text{cut} \lesssim 10^{-3}$, marking a significant improvement in efficiency for these methods.
These results constitute a crucial ingredient for fully differential N$^3$LO calculations based on the $N$-jettiness subtraction scheme.}

\begin{document}
\author[1]{Gherardo Vita}
\emailAdd{gherardo.vita@cern.ch}

\affiliation[1]{CERN, Theoretical Physics Department, CH-1211 Geneva 23, Switzerland}

\maketitle
%
%
\section{Introduction}\label{sec:intro}

As the volume of events collected at the Large Hadron Collider (LHC) and level of refinement in experimental techniques increases, the need for producing more and more precise predictions for LHC observables is becoming of primary importance. Our ability to perform calculations in QCD at higher orders is an essential step in this direction.
While techniques to perform fully differential calculation for hadronic collisions at next-to-next-to-leading order (NNLO) are well established ~\cite{Catani:2007vq, Caola:2017dug,GehrmannDeRidder:2005cm,Czakon:2010td,Boughezal:2011jf,Czakon:2014oma,Boughezal:2015aha,Gaunt:2015pea,DelDuca:2016ily,Dulat:2017brz,Magnea:2018hab,Devoto:2023rpv,Gehrmann:2023dxm}, and analytic results are available for inclusive cross sections at next-to-next-to-next-to-leading (N$^3$LO) for several processes~\cite{Anastasiou:2016cez,Mistlberger:2018etf,Duhr:2019kwi,Baglio:2022wzu}, in the last couple of years a number of results are pushing the state of the art for fully differential calculations of the production of a color singlet towards the N$^3$LO benchmark \cite{Chen:2021isd,Billis:2021ecs,Chen:2021vtu,Chen:2022cgv,Chen:2022xnd,Chen:2022lwc,Neumann:2022lft,Campbell:2023lcy}.
Event shape observables, such as thrust \cite{Farhi:1977sg}, or the $N$-jettiness resolution variable, $\Tau_N$, \cite{Stewart:2009yx, Stewart:2010tn}, can be used to perform subtractions for fixed-order calculations at the LHC \cite{Boughezal:2015aha,Gaunt:2015pea} as well as at $e^+e^-$ colliders. 
For the particular case of color singlet production, there has been intense progress in extending exclusive calculations to N$^3$LO.
With the recent calculation of the three loop $q_T$ beam functions \cite{Luo:2019szz,Ebert:2020yqt}, all the required perturbative ingredients for $q_T$ subtractions \cite{Catani:2007vq} have been computed (the hard function and $q_T$ soft function have been calculated before, see \cite{Gehrmann:2010ue,Li:2016ctv}). 
This has enabled the first predictions for fully differential color singlet production at N$^3$LO accuracy employing the $q_T$ subtraction method \cite{Billis:2021ecs,Chen:2021vtu,Chen:2022cgv,Chen:2022lwc,Neumann:2022lft,Campbell:2023lcy}.
In the last few years, important results have also come out for obtaining the ingredients to extend $N$-jettiness subtraction to N$^3$LO. 
These include the calculation of the hard, jet, beam, and soft function, which are the perturbative ingredients of the leading power factorization theorem for this observable in soft-collinear effective theory (SCET)~\cite{Bauer:2000ew, Bauer:2000yr, Bauer:2001ct, Bauer:2001yt, Bauer:2002nz}.
The N$^3$LO jet function has been calculated in \cite{Bruser:2018rad} and the three-loop beam functions have been obtained in \cite{Ebert:2020unb} (see also \cite{Baranowski:2022vcn}), therefore completely determining the collinear ingredients at this perturbative order.
For the case of 2 light-like directions, which is relevant for color singlet production in hadron collisions, significant advancement has happened for the determination of the three loop soft function \cite{Baranowski:2020xlp,Baranowski:2021gxe,Baranowski:2022khd,Chen:2020dpk}, while the N$^3$LO hard function is the same as for $q_T$ subtraction and has been known for a long time \cite{Gehrmann:2010ue}.
From the scalings of the power corrections \cite{Moult:2016fqy,Moult:2017jsg,Ebert:2018lzn}, it is clear that performing subtractions at this order will dramatically benefit from the knowledge of the power corrections. 
In the recent years, substantial progress has been achieved in understanding the behaviour of QCD amplitudes and cross sections in the soft and collinear limit beyond leading power for colliders, see for example \cite{Moult:2015aoa,Bonocore:2015esa,Bonocore:2016awd,Kolodrubetz:2016uim,Beneke:2017ztn,Beneke:2017mmf,Moult:2017rpl,Chang:2017atu,Feige:2017zci,Beneke:2018rbh,Beneke:2019kgv,Moult:2019mog,Boughezal:2018mvf,Bruser:2018jnc,Moult:2018jjd,Ebert:2018lzn,Beneke:2018gvs,Ebert:2018gsn,Moult:2019uhz,Beneke:2019mua,Buonocore:2019puv,vanBeekveld:2019prq,Beneke:2019oqx,Anastasiou:2020vkr,Laenen:2020nrt,Vita:2020ckn,Liu:2020wbn,Broggio:2021fnr,Liu:2021chn,Liu:2022ajh,Broggio:2023pbu}.
Using the techniques of \cite{Moult:2019uhz}, in this paper we give results for the leading logarithmic (LL) power corrections for both Drell-Yan and gluon-fusion Higgs production for the $0$-jet resolution variable beam thrust~\cite{Stewart:2010tn}. 

We begin with a brief review of $N$-jettiness subtraction~\cite{Boughezal:2015aha,Gaunt:2015pea} and fixing the notation for our calculation. 
$N$-Jettiness subtraction can be used to calculate a cross section $\sigma(\cO)$ for a given $N$-jet process. 
Here $\cO$ is the differential measurement on the final state of the process, and we will leave the dependence on it implicit for the reminder of this section. 
In this work we will focus on the case of color singlet production in hadron collisions, therefore we will work with 0-jettiness.
The 0-jettiness variable ~\cite{Stewart:2010pd, Jouttenus:2011wh} is defined as 
\begin{align} \label{eq:Tau0_0}
 \Tau_0 = \sum_i \min \biggl\{ \frac{2 q_a \cdot k_i}{Q_a} \,, \frac{2 q_b \cdot k_i}{Q_b} \biggr\}
\,,\end{align}
where $k_i$ are the momenta of the hadronic final states.
Here, $q_{a,b}$ are label momenta in SCET, and can be obtained from the mass $Q$ and rapidity $Y$ of the color singlet final state as
\begin{align} \label{eq:Born}
q_a^\mu = x_a \Ecm \frac{n^\mu}{2} = Q e^Y \frac{n^\mu}{2}
\,,\qquad
q_b^\mu = x_b \Ecm \frac{\bn^\mu}{2} = Q e^{-Y} \frac{\bn^\mu}{2}
\,,\end{align}
where we have defined the light-like vectors
\begin{equation}
 n^\mu = (1,0,0,1) \,,\quad \bn^\mu = (1,0,0,-1)
\end{equation}
which are aligned with the proton beam directions, and the Born momentum fractions
\begin{align}
 x_a = \frac{Q e^Y}{\Ecm} \,,\quad x_b = \frac{Q e^{-Y}}{\Ecm}
\,.\end{align}
The normalization factors $Q_{a,b}$ in \eq{Tau0_0} can be used to obtain different versions of the 0-jettiness resolution variable.
Two natural definitions can be obtained by performing the minimization in the hadronic or in the leptonic/color singlet center of mass frame~\cite{Stewart:2009yx, Berger:2010xi}.
This leads to the following definitions
\begin{alignat}{4} \label{eq:Tau0_2}
 &\text{leptonic:}\quad & Q_a &= Q_b = Q \,,\qquad
  & \Tau_0^\lep
  &= \sum_i \min \biggl\{ \frac{x_a \Ecm}{Q} n \cdot k_i \,,\, \frac{x_b \Ecm}{Q} \bn \cdot k_i \biggr\}
\nn\\* &&&&
  &= \sum_i \min \biggl\{ e^Y n \cdot k_i \,,\, e^{-Y} \bn \cdot k_i \biggr\}
\nn\\
 &\text{hadronic:}\qquad & Q_{a,b} &= x_{a,b} \Ecm \,,
 & \Tau_0^\hadcm &= \sum_i \min \Bigl\{ n \cdot k_i \,,\, \bn \cdot k_i \Bigr\}
\,.\end{alignat}
It is by now well understood \cite{Moult:2016fqy,Moult:2017jsg,Ebert:2018lzn} that the hadronic definition suffers from large power corrections that grow exponentially with rapidity.
This can be understood as a manifestation of the fact that the hadronic definition forces an RPI-III~\cite{Manohar:2002fd} (or boost) charge on $\tau$, hence introducing a large dependence on $Y$ in its power expansion coefficients.  
On the other hand, in the leptonic definition, the observable $\tau$ becomes a boost invariant quantity, thanks to the explicit factor of the color singlet rapidity exactly compensating for the RPI charge of the plus or minus component of final state hadronic momenta. 
For this reason, it does not come as a surprise that an expansion in this variable has a much more stable dependence on $Y$.
Given this behavior, modern implementations of $0$-jettiness subtraction utilize the leptonic definition, which will thus also be adopted in the remainder of this paper.

The cross section can be organized as
\begin{align}\label{eq:slicing}
	\sigma &= \int_0^\tauc \df \tau \dsigmadtau  +  \int_\tauc^{\tau_\text{max}} \df \tau \dsigmadtau  
	\nn\\ &= 
	 \int_0^\tauc \df \tau \left[ \frac{\df \sigma^{(0)}}{\df \tau} + \sum_{i>0} \frac{\df \sigma^{(i)}}{\df \tau} \right] + \int_\tauc^{\tau_\text{max}} \df \tau \dsigmadtau 
	\nn\\ &= \sigma_\text{sub}(\tauc) + \Delta\sigma(\tauc) + \int_\tauc^{\tau_\text{max}} \df \tau \dsigmadtau \,.
\end{align}  
For $\tau >0$, by construction of the observable, the process has at least one resolved emission on top of the Born configuration of the process. 
Therefore, the contribution in the region above the cut ($\tau > \tauc > 0$) can be calculated using a lower loop result for a process with an extra jet in the final state. 
All the infrared (IR) singularities related to the Born configuration are completely accounted for by the region below the cut, i.e. $\tau < \tauc$.
In \eq{slicing} we have introduced the notation for the power expansion of the differential distribution in $\tau$ 
\begin{align}
	\frac{\df \sigma}{\df \tau} &=\frac{\df \sigma^{(0)}}{\df \tau} + \sum_{i>0} \frac{\df \sigma^{(i)}}{\df \tau}\,,
\end{align}
where%
\footnote{Note that the notation for  $\frac{\df \sigma^{(i)}}{\df \tau}$ is different compared to the one appearing in \refscite{Moult:2016fqy,Moult:2017jsg,Ebert:2018lzn} where $\frac{\df \sigma^{(i)}}{\df \tau}$ indicates terms suppressed by $\lambda^i$, where $\lambda$ is the SCET power counting parameter, compared to the leading power. For the case of $N$-jettiness we have $\lambda \sim \sqrt{\tau}$.} 
\be
	\frac{\df \sigma^{(i)}}{\df \tau} / \frac{\df \sigma^{(0)}}{\df \tau} \sim \cO\left(\tau^i\right)\,.
\ee
By construction of the $N$-jettiness variable, the leading term of the expansion around $\tau = 0$, $\frac{\df \sigma^{(0)}}{\df \tau}$, contains all the singular terms of the distribution.
The remaining powers in the expansion yield a vanishing contribution for $\tauc\to 0$ after integration. 
Consequently, we have divided the contributions from the region below the cut into two parts: a subtraction term, $\sigma_\text{sub}$, which must be calculated exactly, and a residual term, $\Delta\sigma(\tauc)$, that will be neglected
\be
	\sigma_\text{sub}(\tauc) = \int_0^{\tauc} \df \tau \dsigmadtauat{\text{sub}}\,,\qquad \Delta\sigma(\tauc) = \int_0^{\tauc} \df \tau \left[  \dsigmadtau - \dsigmadtauat{\text{sub}} \right]\,.
\ee
Calculating $\sigma$ as the sum of the subtraction term and the integral above the cut introduces an error due to the omission of the terms in $\Delta\sigma(\tauc)$, which can be minimized by taking $\tauc$ smaller and smaller.
\footnote{A more local way of implementing the subtraction procedure is to act locally in $\tau$ with the subtraction term. 
This can be achieved by adding and subtracting a term as follows
\begin{align}\label{eq:localslicing}
	\sigma &= \int_\tauc^{\tau_\text{max}} \df \tau \left[ \dsigmadtau - \dsigmadtauat{\text{sub}}\theta(\tau < \tau_\text{off})\right] + \int_0^{\tau_\text{off}} \df \tau \dsigmadtauat{\text{sub}}  + \Delta\sigma(\tauc)  \,.
\end{align}
Here $\tau_\text{off}$ determines the region where the subtraction is active.  
Ignoring $\Delta\sigma(\tauc)$ in \eq{localslicing} introduces a $\tauc$-dependent error, similar to \eq{slicing}.
However, the final result remains independent of $\tau_\text{off}$'s value because its contribution is both added and subtracted identically."
The main benefit of \eq{localslicing} is that the cancellation of singular terms between $\dsigmadtau$  and $\dsigmadtauat{\text{sub}}$ happens point by point in $\tau$. 
This in principle allows for improving numerical stability and therefore push $\tauc$ to smaller values, see \refcite{Gaunt:2015pea} for more details.}
The main challenge in the application of $N$-jettiness subtraction is the fact that for the precision requirement of state-of-the-art calculations, very small values of $\tauc$ are necessary. 
This leads to severe numerical instabilities in the calculation of the cross section with an extra emission which, in all practical applications, is obtained via Monte Carlo integration techniques. 
At high perturbative order, and for very small values of $\tauc$, the required level of precision for the above-the-cut result either requires an enormous amount of computing time or it is just beyond the reach of currently available codes.

In what follows, we wish to further investigate the size of $\Delta\sigma(\tauc)$.
In all common implementations of slicing in the literature, the subtraction term only contains the leading power term in the distribution, i.e. $\dsigmadtauat{\text{sub}} =  \frac{\df \sigma^{(0)}}{\df \tau} \sim \frac{1}{\tau}$. 
This implies that $\Delta\sigma(\tauc)$ is the cumulant of the $\tau$ distribution beyond leading power.
Order by order in perturbation theory, we have that these terms take the form
\be \label{eq:NLPscaling}
	\frac{\df \sigma^{(i)}}{\df \tau} \sim \tau^{i-1} \sum_{n=1} \sum_{m=0}^{2n -1} \left(\frac{\alpha_s}{4\pi}\right)^n \ln^m \tau ~c_{n,m}^{(i)}\,,
\ee
with coefficients $c_{n,m}^{(i)}$.
Consequently, at $\cO(\alpha_s^n)$, the residual error naively scales as 
\be \label{eq:deltasigmascalingnaive}
	\Delta\sigma(\tauc) \sim \tauc \sum_{m=0}^{2n -1} \ln^m \tauc + \cO(\taucat{2})\,.
\ee
Hence, at each order in $\alpha_s$, high powers of logarithms are present and they can dramatically enhance the size of these residual uncertainties as one takes $\tauc$ to small values. 
As one approaches $\tauc \to 0$, of course eventually the power suppression will dominate the logarithmic enhancements. 
However, at N$^3$LO ($n=3$), enhanced power corrections up to $\ln^5 \tauc$ demand very low values of $\tauc$ to minimize the impact of $\Delta\sigma(\tauc)$ on the final result.

Before we continue, let us make some technical remarks.
In general, the $c_{n,m}^{(i)}$ are functions of the Born kinematics, and may depend on parton distribution functions in the case of hadron collisions.
The fact that only integer powers of $\tauc$ will appear beyond leading power is not guaranteed.
It is in fact known that, for example, in the presence of fiducial cuts, arbitrary fractional powers of $\tau$ will appear~\cite{Ebert:2019zkb}. 
For example, $p_T$ cuts on the leptonic final states of a color singlet decay will induce power corrections of $\cO(\sqrt{\tauc})$, in analogy to the power corrections of $\cO(|q_T^\text{cut}|)$ appearing in the $q_T$-subtraction formalism when such cuts are applied.
However, these fractional power corrections vanish for sufficiently inclusive observables. 
Moreover, they are purely kinematic in nature. Although they may be difficult to account for analytically in general, we will demonstrate in \Sec{P2BIS}  how to disentangle them from other sources of power corrections by calculating them numerically. 
Therefore, we can disregard these fractional power corrections and consider \eq{deltasigmascalingnaive} as an appropriate starting point for understanding the scaling of the slicing error.
In particular, since the focus of this work is the application of $N$-jettiness subtraction at N$^3$LO, we will focus on the case $n=3$.
\begin{figure*}
 \centering
 \includegraphics[width=0.7\textwidth]{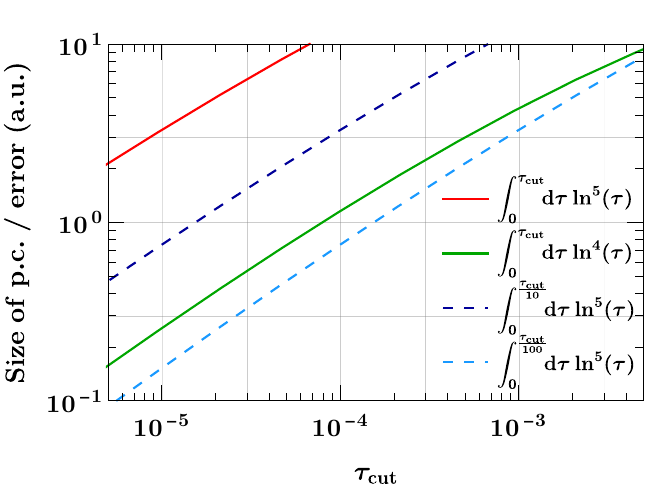}
 \caption{Size of next-to-leading power (NLP) logarithmic contributions to the cumulant cross section. 
 We see that, for this region of $\tauc$, going from a $\ln^5 \tau$ to a $\ln^4 \tau$ dependence for the distribution has a numerical impact comparable to a reduction of $\cO(50)$ in the slicing parameter.
As a consequence, gaining analytic control on single logarithmic powers of the NLP correction may provide improvements on the slicing errors comparable to orders of magnitude of additional computing time.}
 \label{fig:logpowercomparison}
\end{figure*}
In this case,
\be \label{eq:deltasigma2}
	\Delta\sigma^{\mathrm{N3LO}}(\tauc) \sim   \left(\frac{\alpha_s}{4\pi}\right)^3 \int_0^\tauc \df \tau \left( c_{3,5}^\text{NLP}  \ln^5 \tau   +   c_{3,4}^\text{NLP} \ln^4 \tau +   c_{3,3}^\text{NLP} \ln^3 \tau  +  \dots \right)   \,.
\ee
As calculations beyond leading power at N$^3$LO are very challenging, it is interesting to understand what impact may have on the slicing error the calculation of individual terms in the \emph{logarithmic} expansion of the NLP series. 
As a matter of fact, having analytic control on them would allow for their inclusion in the subtraction term $\sigma^\text{sub}(\tauc)$, therefore removing them from contributing to the residual error $\Delta\sigma^{\mathrm{N3LO}}(\tauc)$.
In order to estimate such effect, one may look at the size $\Delta\sigma^{\mathrm{N3LO}}(\tauc)$ with or without the $\ln^5 \tau$ making the assumption that $c_{3,j}^\text{NLP} \sim c_{3,5}^\text{NLP}$ for $j \leq 4$, i.e. that the  numerical coefficients in front of the logarithms are of comparable size%
\footnote{As the coefficients in general depend on several aspects, such as the partonic process, the PDFs, and the color singlet rapidity and invariant mass, there will be cases where the coefficients may differ substantially in size. 
However, for the case of Drell-Yan and gluon-fusion Higgs production at the LHC, assuming a similar size for these coefficients is well in line with the results obtained at NLO and NNLO \cite{Moult:2016fqy,Moult:2017jsg,Ebert:2018lzn}.}.
To get an idea of the size of the numerical improvement due to the analytic control on the LL term, we can compare it to the size of the slicing error $\Delta\sigma^{\mathrm{N3LO}}(\taucat{\prime})$ for $\taucat{\prime} < \tauc $.
In \fig{logpowercomparison}, we show this analysis with some samples values $\taucat{\prime}=\tauc/10, \tauc/100$.
We see that, due to the large power of the logarithm, analytically calculating the leading logarithm at this order, may improve the numerical accuracy of the result by an order of magnitude which translates to obtaining the same accuracy with a cut that is almost 2 order of magnitudes larger than in the case with no power corrections.
Although this is just a qualitative picture, as the real quantitative impact will necessarily depend on the relative size and sign of the coefficients of the logs (which will be functions of the Born kinematic and the partonic channel), the scaling of the improvement is very promising.

In this work we will calculate analytically the leading logarithmic contribution to $\Delta\sigma(\tauc)$ at $\cO(\alpha_s^3)$.

\section{Projection to Born Improved Subtractions}
\label{sec:P2BIS}
In realistic experimental measurements a set of kinematic selection cuts on the final state particles is required.
If these selection cuts are IRC safe, they do not spoil the singular structure of the cross section.
However, they may induce large power corrections to the $\tau$ distribution, which often times are significantly bigger than the ones due to subleading power QCD dynamics.
In the case of $q_T$ subtraction for color singlet production, a boost prescription can be implemented to capture these fiducial power corrections at the level of the factorization theorem \cite{Catani:2015vma,Ebert:2019zkb,Ebert:2020dfc,Camarda:2021jsw,Buonocore:2021tke}, but this method fails in the presence of photon isolation cuts.
Unfortunately, in the more general cases, accounting for these cuts analytically in the subtraction term is very challenging, since they consist of non-analytic constraints on the phase space of particles.
Here, following a method proposed in \refcite{Ebert:2019zkb}, we will show how one can separate fiducial power corrections from other sources of power corrections%
\footnote{We will refer to power correction not related to fiducial cuts as \emph{dynamical} power corrections as they are due to correction to the leading power dynamics of QCD}
and obtain them numerically.
For concreteness, we focus on the case of Higgs production in gluon fusion at N$^3$LO with di-photon decay. 
Given the desired final state measurement $\Obs$, which we allow for including fiducial cuts, instead of computing the cross section $\sigma(\Obs)$, we use $0$-jettiness subtraction to compute the cross section for another observable $\sigma(\Obst)$, with $\Obst$ being a Born projection of the measurement $\Obs$. 
The difference $\sigma(\Obs) - \sigma(\Obst)$  vanishes for $\tau \to 0$ for any infrared safe observable $\cO$.
This implies that $\sigma(\Obs) - \sigma(\Obst)$ can be calculated using only information about the NNLO Higgs + jet process.
In formulas this takes the form
\begin{align}
    \sigma_{h,\,\mathrm{N^3LO}}(\Obs) &= \sigma_{h,\,\mathrm{N^3LO}}(\Obst) + \sigma_{h+j,\,\mathrm{NNLO}}(\Obs - \Obst) \\[.5cm]
    &= \int^\tauc_0 \df \tau \frac{\df \sigma^\text{sub}_{h,\,\mathrm{N^3LO}}}{\df \tau}(\Obst) + \int_{\tau > \tauc} \df \sigma^\text{full}_{h+j,\,\mathrm{NNLO}}(\Obst) &\text{\small(Slicing for $\Obst$)} \nonumber\\[.5cm]
    &\quad+  \int^\tauc_0 \df \tau \left[\frac{\df \sigma^\text{full}_{h,\,\mathrm{N^3LO}}}{\df \tau} - \frac{\df \sigma^\text{sub}_{h,\,\mathrm{N^3LO}}}{\df \tau}\right](\Obst)    &\text{\small(Slicing residual for $\Obst$ at $\tau < \tauc$ )} \nonumber\\[.5cm]
    &\quad+   \int \df \sigma^\text{full}_{h+j,\,\mathrm{NNLO}}(\Obs - \Obst)    &\text{\small (P2B correction factor for $\Obs$ vs $\Obst$)}\nn
\end{align}
It is easy to recognize that the Projection to Born (P2B) correction factor precisely corresponds to the subtracted real emission term of a P2B subtraction in $pp$ \cite{Cacciari:2015jma,Chen:2021isd}, with all the nice numerical features that P2B subtraction carries.
The main advantage of this P2B improved slicing from the point of view of the power corrections is that the slicing calculation is performed for a Born observable. 
This means that the residual corrections of the slicing only depend on Born kinematics and therefore they are insensitive to effects of the fiducial cuts on the full phase space (which are the ones responsible for the presence of large fiducial power corrections).
The physical effect of the presence of the fiducial cuts is accounted for by the P2B correction factor that enjoys the numerical stability of the P2B subtraction, namely the fact that, point by point in the phase space of the real radiation, the counterterm is given by the full real matrix element. 
Notice that using projection to Born improved subtractions would not change the fact that these cuts may severely affect the perturbative convergence of the fixed order distributions in certain regions \cite{Chen:2021isd,Billis:2021ecs}, as this is an issue of fixed order predictions and not of the way those are obtained.
Moreover, even in the case of different fiducial cuts, such as the \emph{quadratic}-cuts from \refcite{Salam:2021tbm}, one would still benefit from employing the P2B improved subtraction compared to the traditional one. 
This is the case because the main effect of those cuts is to remove the \emph{linear} power corrections (which in the case of $0$-jettiness would be the ones scaling as $\cO(\sqrt{\tauc})$), but they would modify the power corrections beyond that.
Therefore, in order to analytically improve the subtraction without employing the P2B improved subtraction, one would need to analytically calculate both the dynamical power corrections as well as the fiducial ones, now entering at $\cO(\tauc)$.

\section{Leading Logarithmic Power Corrections at N$^3$LO}
In this section we give the results for the leading logarithmic power corrections for color singlet production at the LHC at N$^3$LO using 0-jettiness subtraction.
We start by using consistency relations to understand what is the minimal set of ingredients needed to obtain the result.
We will then present the results for the case of Drell-Yan and Higgs production in gluon fusion.
\subsection{Consistency Relations and Differential Color Singlet Production Beyond Leading Power}
The cancellation of IR singularities order by order in perturbation theory imposes constraints on the structure of the differential distribution in $\tau$. 
We can leverage these constraints to derive consistency conditions between different contributions entering at NLP \cite{Moult:2016fqy}.
Here, we work out the consistency conditions to N$^3$LO.
The modes necessary to describe the process to the order we are working are hard, (ultra)soft, and collinear.
Note that we do not distinguish between $n$ and $\bn$ collinear modes in this counting as they will behave identically for the purposes of this section. 
In the same spirit we will use the word \emph{loop} loosely, indicating both loop corrections as well as phase space integrals, since, mode by mode, they both contribute in dimensional regularization with the same power in the $\epsilon$ expansion to the cross section.
Beyond leading power, we always need at least one collinear or soft mode, since otherwise we are confined to the Born configuration whose contributions are always proportional to $\delta(\tau)$, which clearly constitutes a leading power term. 
The next-to-leading-power term of the differential distribution in $\tau$ at bare level takes the form
\begin{align}\label{eq:mastereqconsistency}
	\frac{\df \sigma^{\NLP}}{\df \tau} = \sum_n \left(\frac{\alpha_s}{4\pi}\right)^n \times \left[ \sum_{t=0}^{n-1} \sum_{\kappa \in \mathfrak{M}_{n-t}} \sum_{i=0}^{2n-1} \frac{c^t_{k,i}}{\epsilon^i}\left(\frac{\mu^2}{Q^2 \tau^{\frac{m(\kappa)}{(n-t)}}}\right)^{(n-t)\epsilon}\right]\,,
\end{align}
where $\mathfrak{M}_{l}$ is the set of all possible sets of modes of length $l$ excluding purely hard contributions, $m(\kappa)$ is the scaling of the list of modes $\kappa$ and $t>0$ represents terms that will be removed by the interference of UV counterterms (in the full theory) with lower order results, such as coupling renormalization terms.
To make this definitions more concrete, at $\cO(\alpha_s^3)$, we have contributions involving up to 3 modes, so
\be
	\mathfrak{M}_3 = \Big\{ \{s,s,s\},\{c,c,c\},\{c,c,s\},\{h,h,c\},\{h,c,s\},\dots \Big\}\,, \qquad \{h,h,h\} \notin \mathfrak{M}_3\,.
\ee
The $\tau^{\epsilon m(\kappa) }$ scaling for a set of modes is determined by the characteristic scales associated with the soft modes $\mu_s^2 \sim Q^2 \tau^2$, collinear modes $\mu_c^2 \sim Q^2 \tau$, and the hard modes $\mu^2_h \sim Q^2$.
To give some examples of scaling at three loops:
\begin{align}
	m(\{s,s,s\}) &= 6\,,\quad m(\{c,s,s\}) = 5 \,,\quad m(\{c,c,s\}) = 4\,, \nn\\
	m(\{h,c,s\}) &= 3 \,,\quad m(\{h,h,s\}) = 2\,,\quad m(\{h,h,c\}) = 1 \,.
\end{align}
Due to the requirement of having at least one $c$ or $s$ mode contributing at NLP, we always have $m(\kappa)>0$.
Equipped with this notation, we can expand \eq{mastereqconsistency} at each order in $\alpha_s$ up to $\cO(\epsilon^0)$ and impose the cancellation of all poles.
We work out the constraints at N$^3$LO and we write the N$^3$LO distribution at NLP in terms of the minimal number of coefficients
\begin{align}\label{eq:taudistrafterconsrel}
	\frac{1}{\sigma_0}\frac{\df \sigma^{\NLP}}{\df \tau}\Big|_{\cO(\alpha_s^3)} =&~ c_{hhc,5}\ln^5 \tau 
	\\ 
	&+(c_{hhc, 4} + c_{sss,4} - d_{ss,4}) \ln^4 \tau + d_{ss,4} \ln\frac{\mu^2}{Q^2}  \ln^3 \tau  \nn\\
	&+(4 c_{hhc, 3} + c_{hhs,3} + c_{hcc,3}  -  c_{sss, 3} + 2 d_{hc,3} + d_{ss,3} - e_{s, 3})\ln^3 \tau  + \dots \nn
\end{align}
where the dots include terms with lower powers of $\ln \tau$ and constants, and for ease of notation we used $c^0_{k,i} \equiv c_{k,i},~c^1_{k,i} \equiv d_{k,i}, ~c^2_{k,i} \equiv e_{k,i}$ . 
From \eq{taudistrafterconsrel} we see that for the coefficient of the leading log, $\ln^5 \tau$, one only needs to calculate the leading, $1/\epsilon^5$, pole of the contribution coming from two hard loops and a single collinear emission.

For this reason, we consider the production of a color singlet $h$ in proton-proton collisions in association with a single collinear emission of momentum $k$
\be
	p(p_a) + p(p_b) \to h(q) + k  \,.
\ee
We want to retain all information on the Born kinematics of the color-singlet such that the calculation of our power corrections can be applied to fully differential calculations.
Following the common practice, we parametrize the Born phase space using the rapidity and the invariant mass of the color singlet
\be
	Q^2 = q^2 \,,\qquad Y= \frac{1}{2}\ln\left(\frac{p_b \cdot q}{p_a \cdot q}\right)\,.
\ee
The contribution from a single collinear emission along the $n$-direction at NLP in the $0$-jettiness distribution, differential in the color singlet rapidity and invariant mass and using the leptonic definition of 0-jettiness, reads \cite{Ebert:2018lzn}
\begin{align} \label{eq:sigma_NLO_NLP_coll_LL}
 \frac{\df\sigma_n^{\NLP}}{\df Q^2 \df Y \df\Tau} &
 = \int_{x_a}^1 \frac{\df z_a}{z_a} \,
   \frac{Q}{2 x_a x_b \Ecm^4}
   \frac{(Q^2 \tau/\mu^2)^{-\eps}}{(4\pi)^2(1-z_a)^\eps} \biggl\{
   f_a\!\left(\!\frac{x_a}{z_a}\!\right) f_b(x_b) \, \Msquared^{(2)}(k[\tau,z_a],\epsilon)
   \nn\\&\qquad
	\tau \Msquared^{(0)}(k[\tau,z_a],\epsilon) \biggl[
	    -f_a\!\left(\!\frac{x_a}{z_a}\!\right) f_b(x_b)\, + \, f_a\!\left(\!\frac{x_a}{z_a}\!\right) x_b  f'_b(x_b) \biggr]
   \biggr\}
\,,\end{align}
where $k$ is fixed by the measurement constraints to be
\begin{align} \label{eq:k_collinear}
 k^\mu [\tau,z_a] = x_a \Ecm \frac{1-z_a}{z_a} \frac{n^\mu}{2} + \frac{\tau \Ecm}{x_a} \frac{\bn^\mu}{2}
       + Q\sqrt{\tau}\sqrt{\frac{1-z_a}{z_a}} n_\perp^\mu
\,,\end{align}
and $ \Msquared^{(2)} \sim \tau \Msquared^{(0)} $ is the subleading power expansion of the matrix element.
In \eq{sigma_NLO_NLP_coll_LL} we see that we get two contributions: 
The first term accounts for the contribution from subleading power matrix elements integrated over the leading power collinear phase space.
The second term in the braces corresponds to the terms involving the leading power matrix element integrated over the subleading power expansion of the phase space, which in this case includes the Born measurement functions defining $Q$ and $Y$ as well as $\tau$ in terms of the momenta $p_a,p_b$, and $k$ of the color charged particles.
The leading power matrix elements can be obtained using two loop splitting amplitudes, while the amplitudes at NLP for a single collinear emission at two loops where derived in \refcite{Moult:2018jjd,Moult:2019uhz} from the full kinematic results of \refcite{Gehrmann:2011aa,Garland:2001tf,Garland:2002ak}.

\begin{figure*}
 \centering
 \includegraphics[width=0.4\textwidth]{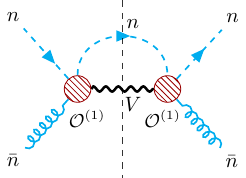}
 \hfill
 \includegraphics[width=0.4\textwidth]{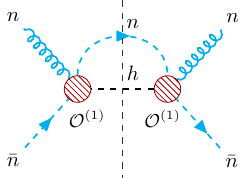}
 \caption{Example SCET cut diagrams for the off-diagonal channels contributing to leading logarithm at next-to-leading-power. 
 In blue are depicted collinear fields, while in black the color singlet particles.
 The red blobs represent subleading hard scattering operators whose Wilson coefficients will encode the virtual loop corrections to the process. 
 For our N$^3$LO analysis, one needs to consider the square of the 1-loop Wilson coefficients and the interference of the 2-loop and tree-level Wilson coefficients.}
 \label{fig:off_diagonal_plots}
\end{figure*}

The less obvious results are the ones for the off-diagonal subleading power matrix element related to the soft quark Sudakov derived in \refcite{Moult:2019uhz} and we report them here for completeness.
The one for Higgs production reads
\begin{align}\label{eq:A2gqHiggs}
	A_{gq,H}^{(2),LL}(k,\epsilon) =  \Msquared^\LO_{gg\to H}(Q)  \times \frac{8 \pi \as C_F }{(1-z_a) z_a} \frac{1}{2}\left[\left(\frac{\alpha_s}{4\pi}\right)\frac{4 C_F + 4 (C_A - C_F) (1-z_a)^{-\epsilon}}{\epsilon^2}\right]^2\,,
\end{align}
while for Drell-Yan we have
\begin{align}\label{eq:A2qgDY}
	A_{qg,DY}^{(2),LL}(k,\epsilon) =  \Msquared^\LO_{q\bar{q}\to V}(Q)  \times 4 \pi \as T_F  \left(\frac{2+z_a^2-2z_a}{1-z_a}\right)\left[\left(\frac{\alpha_s}{4\pi}\right)\frac{4 C_F + 4 (C_A - C_F) (1-z_a)^{-\epsilon}}{\epsilon^2}\right]^2\,.
\end{align}
These subleading power matrix elements have no $\tau$ dependence since they scale as $\tau^0$, which is suppressed compared the $1/\tau$ behavior of the matrix element squared at LP.
Notice the divergent $z_a \to 1$ behaviour of these subleading power matrix elements which is responsible for promoting to leading logarithm the contributions of these off-diagonal channels.

\subsection{Analytic Results}
We now present the explicit results for the leading logarithmic correction for 0-jettiness at N$^3$LO for Drell-Yan and Higgs production.
In both cases we can factor out the leading order partonic cross section 
\be
	\hat{\sigma}^\text{LO}(Q) = \frac{|\cM^\text{LO}|^2}{2 x_a x_b \Ecm^4} = 
	\begin{cases}
		\frac{4\pi \alpha_{em}^2}{3 N_c Q^2 \Ecm^2}
 \biggl[ Q_q^2 + \frac{(v_q^2 + a_q^2)(v_l^2 + a_l^2) - 2 Q_q v_q v_l (1 - m_Z^2/Q^2)}{(1-m_Z^2/Q^2)^2
         + m_Z^2 \Gamma_Z^2 / Q^4} \biggr] &\text{Drell-Yan} \\
		\frac{2\pi \delta(Q^2 - m_H^2)}{2 Q^2 \Ecm^2}\frac{\as^2 Q^4}{576 \pi^2 v^2} &\text{Higgs,}
	\end{cases}
\ee
and write the cross section at NLP as 
\begin{align} \label{eq:sigma_NLP}
\frac{\df\sigma^{\text{NLP}(n)}}{\df Q^2 \df Y \df \tau}
&= \hat\sigma^\LO(Q) \, \Bigl(\frac{\as}{4\pi}\Bigr)^n
    \int_{x_a}^1 \frac{\df z_a}{z_a} \int_{x_b}^1 \frac{\df z_b}{z_b}
\biggl[
f_i\biggl(\frac{x_a}{z_a}\biggr) f_j\biggl(\frac{x_b}{z_b}\biggr) C_{a_s^n f_i f_j}^{\text{NLP}}(z_a, z_b, \tau)
\\\nn & \quad
+ \frac{x_a}{z_a} f'_i\biggl(\frac{x_a}{z_a}\biggr) f_j\biggl(\frac{x_b}{z_b}\biggr) C_{a_s^n f_i' f_j}^{\text{NLP}}(z_a, z_b, \tau)
+ f_i\biggl(\frac{x_a}{z_a}\biggr) \frac{x_b}{z_b} f'_j\biggl(\frac{x_b}{z_b}\biggr) C_{a_s^n f_i f_j'}^{\text{NLP}}(z_a, z_b, \tau)
\biggr]
\,.\end{align}

\subsubsection{Drell-Yan}
\label{sec:DY_analytic}
For the case of Drell-Yan, we find for the diagonal channel
\begin{align}\label{eq:DYqqbar}
	C_{a_s^3 f_q f_{\bar{q}}}^{\text{NLP}}(z_a, z_b, \tau) &= 64 C_F^3 \log^5(\tau) \delta(1-z_a)\delta(1-z_b) \,,\nn\\
	C_{a_s^3 f^{\prime}_q f_{\bar{q}}}^{\text{NLP}}(z_a, z_b, \tau) &= C_{a_s^3 f_q f^{\prime}_{\bar{q}}}^{\text{NLP}}(z_a, z_b, \tau) = -\frac{1}{2}C_{a_s^3 f_q f_{\bar{q}}}^{\text{NLP}}(z_a, z_b, \tau)\,.
\end{align}
The fact that the result is proportional to $\delta$-functions in $\{z_a,z_b\}$ is expected due to the soft-collinear consistency of the leading logarithmic term.
As a matter of fact, hard and soft modes are confined to $z_a = z_b = 1$, therefore every soft pole of hard and soft nature will be accompanied by a corresponding delta function in these splitting variables.
In this work we obtained the results from the collinear limit, which in general has the full phase space available in $z_{a,b}$, however, since the leading divergence will necessarily need to cancel against the soft one, the leading logarithmic contribution needs to be proportional to $\delta(1-z_a)\delta(1-z_b)$.

At NLP the leading logarithmic correction receives contributions also from the off-diagonal channel $qg$. 
For this channel we obtain
\begin{align}
	C_{a_s^3 f_q f_{g}}^{\text{NLP}}(z_a, z_b, \tau) &=-\frac{16}{3}T_F (C_A^2 + C_A C_F + C_F^2)  \log^5(\tau)\delta(1-z_a)\delta(1-z_b)\nn\\
	C_{a_s^3 f^{\prime}_q f_{g}}^{\text{NLP}}(z_a, z_b, \tau) &= C_{a_s^3 f_q f^{\prime}_{g}}^{\text{NLP}}(z_a, z_b, \tau) = 0 
\end{align}
These leading logarithmic off-diagonal contributions are generate by subleading power matrix elements and encode contributions due to soft quark emissions and subleading power quark-gluon splitting functions, for details see \cite{Moult:2019uhz}.
Note again the presence of delta functions in $z_{a,b}$. 
This may seem surprising at first, since at leading power off-diagonal channels do not have contributions from soft kinematics, but that is not the case beyond leading power.

These results encode the full LL contributions at NLP at this order. 
The other channels start to contribute only beyond the LL at NLP.

\subsubsection{Higgs Production in Gluon Fusion}
\label{sec:ggH_analytic}
For the case of Higgs in gluon fusion we get leading logarithmic corrections to both the diagonal channel $gg$ as well as for $gq$. 
At NLP for the $gg$ channel, there are contributions both from the expansion of the phase space as well as from the matrix element.
Combining both together we get 
\begin{align}\label{eq:Higgsgg}
	C_{a_s^3 f_g f_g}^{\text{NLP}}(z_a, z_b, \tau) &= 64 C_A^3 \log^5(\tau) \delta(1-z_a)\delta(1-z_b) \,,\nn\\
	C_{a_s^3 f^{\prime}_g f_g}^{\text{NLP}}(z_a, z_b, \tau) &= C_{a_s^3 f_g f^{\prime}_g}^{\text{NLP}}(z_a, z_b, \tau) = -\frac{1}{2}C_{a_s^3 f_g f_g}^{\text{NLP}}(z_a, z_b, \tau)\,.
\end{align}
As for the case of Drell-Yan we get PDF derivatives for the diagonal channel. 
This is due to the kinematic expansion of the momentum fraction beyond leading power and it is one of the contributions coming from the subleading power expansion of the phase space.

We also have a leading logarithmic contribution for the off-diagonal channel $gq$. 
At LP, this channel does not contribute to the LL since the matrix element is not singular in the soft and collinear limit. 
At NLP, there is a singularity associated with the emitted quark going soft and collinear which promotes this contribution to a leading logarithm~\cite{Moult:2017rpl,Moult:2019uhz}. 
This can be clearly seen from the $z_a \to 1$ behaviour of \eq{A2gqHiggs}.
Therefore, the leading logarithmic contribution for this channel comes from the singular behaviour of this subleading power amplitude, while the expansion of the phase space to subleading power only generates subleading logarithmic contributions.
For this channel we get
\begin{align}
	C_{a_s^3 f_g f_q}^{\text{NLP}}(z_a, z_b, \tau) &=-\frac{16}{3}C_F (C_A^2 + C_A C_F + C_F^2)  \log^5(\tau)\delta(1-z_a)\delta(1-z_b)\,,\nn\\
	C_{a_s^3 f^{\prime}_g f_q}^{\text{NLP}}(z_a, z_b, \tau) &= C_{a_s^3 f_g f^{\prime}_q}^{\text{NLP}}(z_a, z_b, \tau) = 0 \,.
\end{align}
The absence of PDF derivatives in this channel at this order is expected and it is one manifestation of the lack of contributions due to subleading power phase space expansion discussed above.
Note that in general, beyond leading power, one may encounter PDF derivatives even integrating a subleading power matrix element against the leading power phase space.
This is due to the fact that subleading power matrix elements may develop power law divergences \cite{Ebert:2018gsn,Bhattacharya:2018vph,Boughezal:2019ggi}. 
In dimensional regularization these divergences generate higher-order plus distributions which involve derivatives of delta functions or plus distributions acting on the leading power phase space and therefore on the collinear PDFs in the case of hadronic collisions.
For beam thrust this is not the case at NLP, however note that for $p_T$ distributions beyond leading power, and therefore for power corrections needed for improving $q_T$ subtraction, this indeed happens, see \refcite{Ebert:2018gsn} for details.

\section{Numerical Results}\label{sec:numerics}
In this section we analyze the numerical size of the power corrections calculated in the previous sections. 
For concreteness, we will focus on Higgs production in gluon fusion in the infinite top mass approximation and Drell-Yan in proton-proton collision. 
We will consider a collider center of mass energy $E_{cm} = 13.6~\mathrm{TeV}$ and employ {\texttt{PDF4LHC15\_nnlo\_mc}} PDFs, with $\alpha_s(m_Z) = 0.118$. 
We do not expect any of the conclusions in this section to depend in significant way from these choices.
Following the discussion in \Sec{intro} the residual slicing error, in the usual setup where the subtraction term consists of only the leading power distribution, can be written as
\begin{align}\label{eq:deltasigma3}
	\Delta \sigma^{N^3LO} (\tauc,Y,Q) &= \int_0^{\tauc} \df \tau \left[  \frac{\df \sigma}{\df \tau \df Q \df Y} - \frac{\df \sigma^\text{LP}}{\df \tau \df Q \df Y} \right]_{\cO(\alpha_s^3)} 
	\\&=
	\left(\frac{\alpha_s}{4\pi}\right)^3 \int_0^\tauc \df \tau \Big( c_{3,5}^\text{NLP}(Q,Y)  \ln^5 \tau   +   c_{3,4}^\text{NLP}(Q,Y) \ln^4 \tau  
	 \nn\\ &\qquad~~\qquad\qquad\qquad+  
	 c_{3,3}^\text{NLP}(Q,Y) \ln^3 \tau  +  \dots \Big) \nn
\end{align}
In this section we will present the results for the cumulant of the power corrections calculated in \secs{DY_analytic}{ggH_analytic}, differential in rapidity, and $Q^2$. 
This shows the numerical impact of our analytic calculation and it is a proxy for $\Delta \sigma^{N^3LO} (\tauc,Y,Q)$ since, for $\tauc \ll 10^{-2}$, the leading logarithmic term is expected to be the dominant contribution.
For convenience, from now on we will fix $Q=91.2~\mathrm{GeV}$ for Drell-Yan and $Q=125.5~\mathrm{GeV}$ for Higgs production and drop the dependence on $Q$ for the rest of the discussion in this section.
As $\Delta \sigma^{N^3LO} (\tauc,Y)$ strongly depends on $\tauc$, we will present results for three reference values: $\tauc = \{ 10^{-3},10^{-4},10^{-5} \}$.

\subsection{Drell-Yan}\label{sec:DY_numerics}

\begin{figure*}
 \centering
 \includegraphics[width=0.8\textwidth]{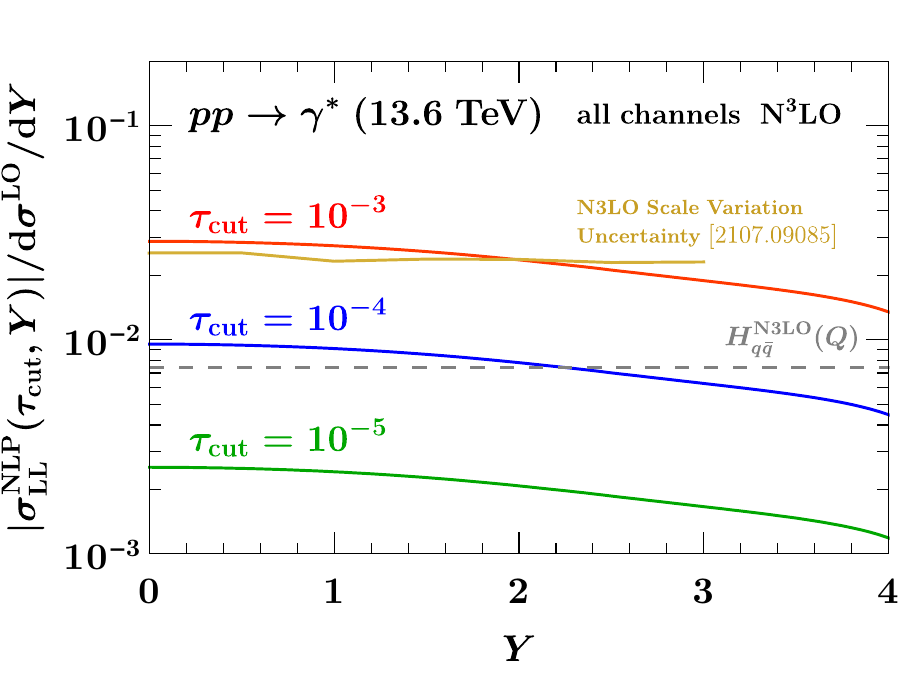}
 \caption{Size of the leading logarithmic power corrections at N$^3$LO for Drell-Yan differential in rapidity. 
 This result accounts for all the contributions entering at leading logarithmic accuracy, which at next-to-leading power include the off-diagonal channel $qg$ due to contributions of soft quarks and fermion number changing collinear splitting. For comparison, the scale variation uncertainties of the N$^3$LO spectrum from \cite{Chen:2021vtu} and the N$^3$LO hard function contribution to the subraction term are plotted, showing that these power corrections are sizable for $\tauc \gtrsim 10^{-5}$.}
 \label{fig:DY_all_channels}
\end{figure*}

In \figs{DY_all_channels}{DYchannel_decomposition} we show the cumulant of the power corrections calculated in \Sec{DY_analytic}, normalized by the leading order rapidity spectrum for different initial state channels.
To get a quantitative understanding of the size of such NLP terms, we compare them to two interesting quantities. 
In dark yellow is the scale uncertainty of the N$^3$LO Drell-Yan rapidity spectrum from \refcite{Chen:2021vtu}. 
This provides a reference for the perturbative uncertainty of this class of processes at N$^3$LO and one would like the numerical error of the N$^3$LO correction, including the slicing one, to be significantly smaller compared to such uncertainty.
The other quantity we decided to plot for numerical comparison is the size of the three loop hard function boundary $H_{q\bar{q}}(Q)$ \cite{Gehrmann:2010ue}, which is one of the N$^3$LO corrections to the singular distribution and it is shown in dashed gray. 
Note that $H_{q\bar{q}}(Q)$ enters the singular distribution with a $\delta(\tau)$. As we are comparing objects at the cumulant level, it is therefore important to point out that its functional dependence on $\tau$ always integrates to 1 irrespectively of the value of $\tauc$. 
Overall, $H_{q\bar{q}}(Q)$ gives a rough idea of the size of the boundary ingredients that enter in the subtraction term at this order.
\begin{figure*}
 \centering
 \includegraphics[width=0.49\textwidth]{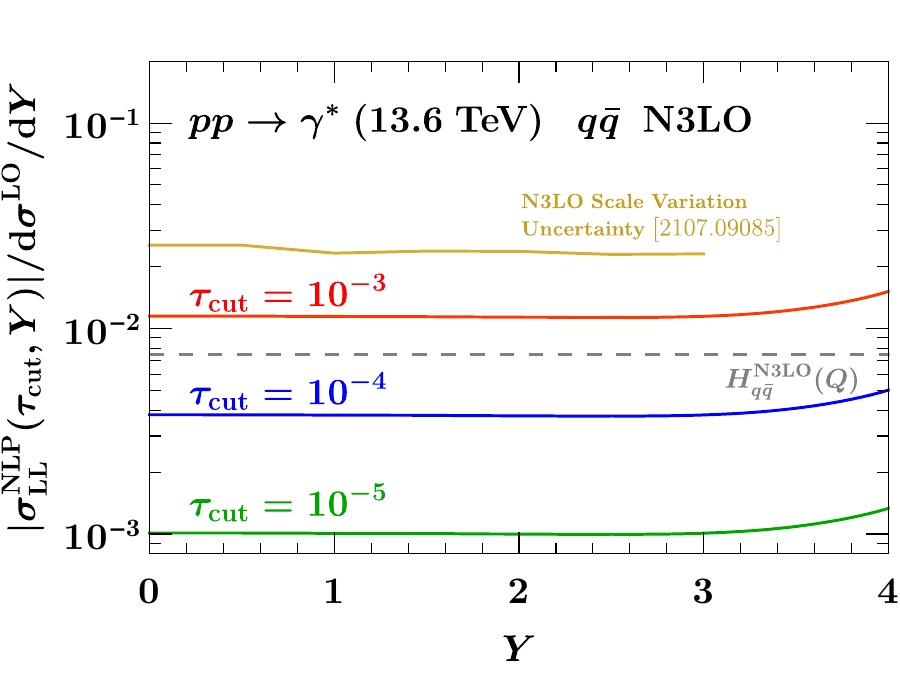}
 \hfill
 \includegraphics[width=0.49\textwidth]{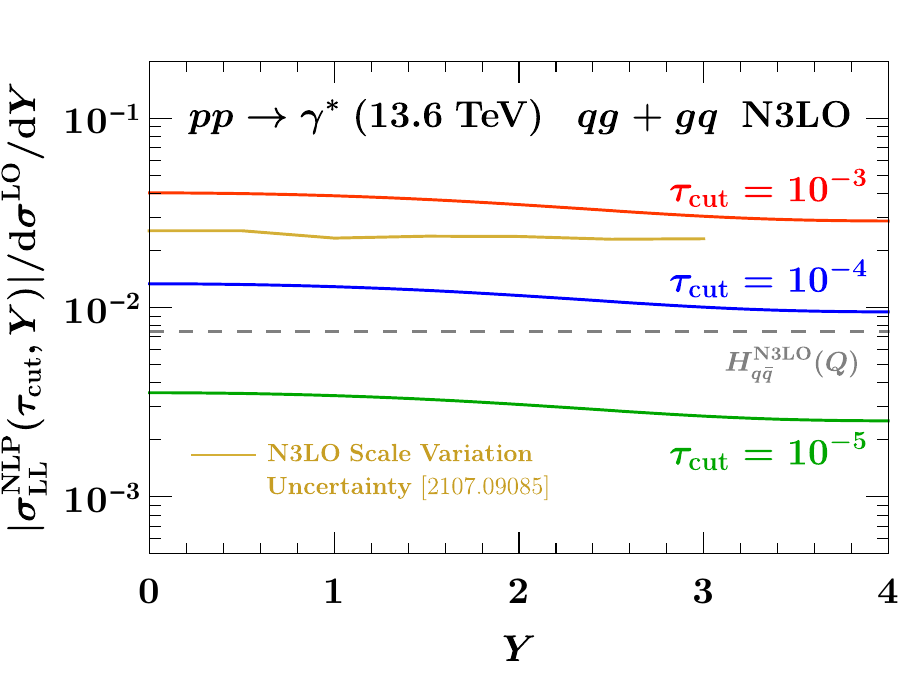}
 \caption{Same as \fig{DY_all_channels} but for different production channels. 
 We show the diagonal channel $q\bar{q}$ on the left and $qg$ on the right. 
 Both channels contribute at leading log at NLP. 
 It is evident that the power corrections to $qg$ are substantially larger than those for $q\bar{q}$.
 Note that the plot shows the absolute value of the LL NLP cumulant, however there is a relative sign between these two channels.}
 \label{fig:DYchannel_decomposition}
\end{figure*}
In \fig{DY_all_channels} we present the result for the sum of all the channels while in \fig{DYchannel_decomposition} we provide a breakdown of the contributions for the two production channels that enter at LL NLP.
We see that the LL NLP term gives contributions at the order of percent for $\tauc = 10^{-3}$ and starts to become negligible compared to the size of N$^3$LO scale variation uncertainties and hard function contribution only starting from $\tauc = 10^{-5}$. 
It is interesting to notice in the channel by channel comparison of \fig{DYchannel_decomposition}, that the power corrections coming from the off-diagonal $qg$ channel are noticeably larger than the ones for the diagonal $q\bar{q}$ channel. 
The fact that for Drell-Yan at N$^3$LO large corrections beyond leading power are coming from the $qg$ channel is not completely surprising. 
As a matter of fact, in the applications of $q_T$-subtraction to these processes it was empirically observed that the most challenging channel for taming the effect of NLP terms was indeed the $qg$ channel~\cite{Chen:2021vtu}.
On the one hand, this is due to the fact that at LHC energies there is an abundance of gluons from the PDFs, but it is important to recall that at NLP the off-diagonal $qg$ channel is not logarithmically suppressed.
Although our calculation focuses only on $0$-jettiness and therefore it cannot be considered an analytic demonstration of the behavior seen in the applications of $q_T$-slicing, the feature of the PDF enhancement and that this off-diagonal channel will contribute at leading log beyond leading power, we expect to generalize to the case of $q_T$-slicing very naturally, since it was already shown analytically at lower order \cite{Ebert:2018gsn}. 

\begin{figure*}
 \centering
 \includegraphics[width=0.8\textwidth]{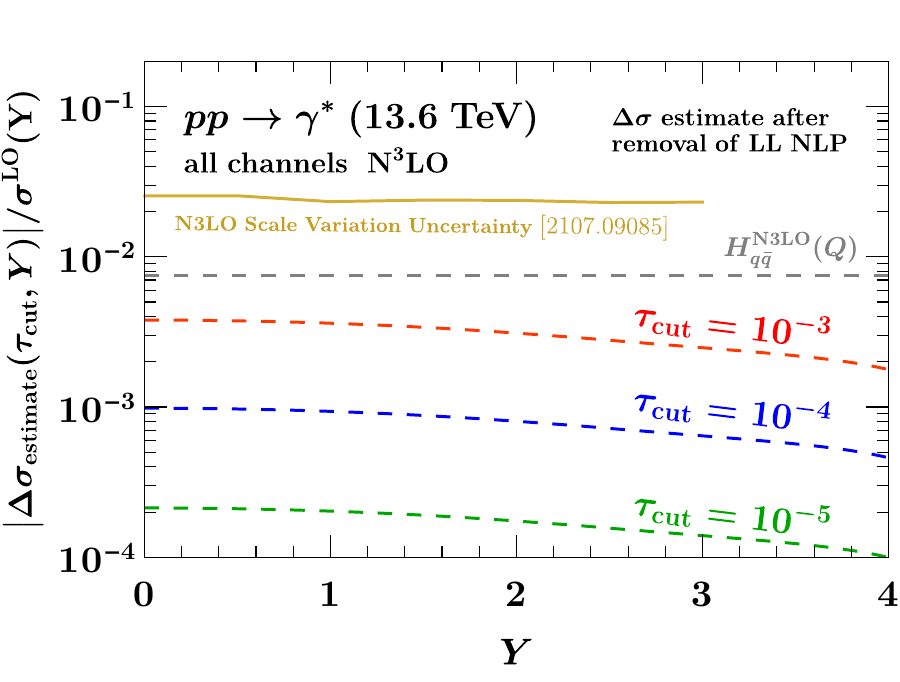}
 \caption{Estimate of the slicing error $\Delta \sigma(\tauc,Y)$ after the removal of the leading logarithmic terms analytically calculated in this work (i.e. after their inclusion in the subtraction term) for various values of $\tauc$. 
 The result is normalized by the leading order rapidity spectrum and plotted for comparisons are the scale variation uncertainties of the N$^3$LO spectrum from \cite{Chen:2021vtu} and the N$^3$LO hard function contribution to the subraction term.} 
 \label{fig:DYresiduals}
\end{figure*}

Given the experience at NNLO, where already tens of thousands of CPU hours are necessary to obtain reliable NLO above-the-cut results for $\tauc < 10^{-3}$, see for example \refscite{Alekhin:2021xcu,Campbell:2022gdq}, it seems unrealistic to expect that current NNLO codes for color singlet plus jet will be able to push to $\tauc \sim 10^{-5}$ without requiring an outrageous amount of computing resources.
Hence, these power corrections will have significant impact for realistic applications of 0-jettiness at N$^3$LO.
In this regard, in \fig{DYresiduals} we try the exercise of estimating the size of the slicing error if the leading logarithmic corrections at NLP are included in the subtraction term.
Clearly, the real size of the residual error will ultimately depend on the actual numerical coefficients of the subleading logarithmic (and power) terms, so some assumptions are necessary for this estimate.
In the following we look at the case where $c_{3,4}^\text{NLP}(Y) = c_{3,5}^\text{NLP}(Y)$, i.e. we use for the next-to-leading logarithmic coefficient the same numerical value we obtained for the leading log. 
We neglect further subleading logs and powers as we expect that for such small values of $\tauc$ they will not play a significant role.

As we can observe from \fig{DYresiduals}, in this situation the power corrections would be noticeably smaller than the scale variation already starting from $\tauc \sim 10^{-3}$ and completely negligible for $\tauc < 10^{-4}$. 
Being able to keep the slicing error under control while loosening the cuts so significantly would be a very important improvement in the feasibility and efficiency of slicing calculations at this order.

\subsection{Higgs Production in Gluon Fusion}\label{sec:Hgg_numerics}
We now turn our attention to the case of Higgs production in gluon fusion.
\begin{figure*}
 \centering
 \includegraphics[width=0.8\textwidth]{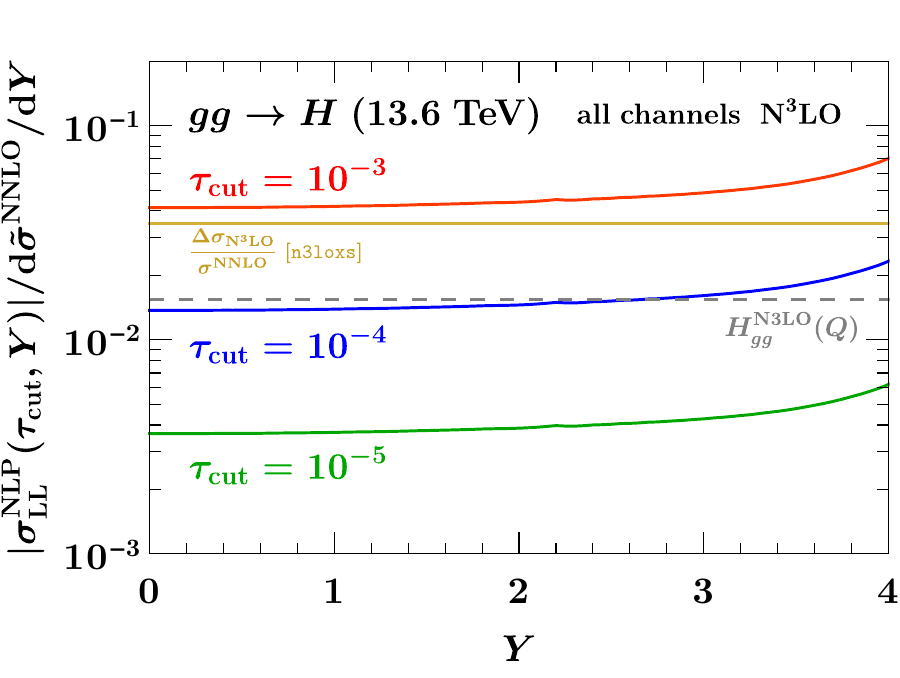}
 \caption{Size of the leading logarithmic power corrections at N$^3$LO for Higgs production in gluon fusion differential in rapidity. 
 This result accounts for all the contributions entering at leading logarithmic accuracy, which at next-to-leading power include the off-diagonal channel $gq$ due to contributions of soft quarks and fermion number changing collinear splitting. 
 For comparison, the N$^3$LO inclusive K-factor from \cite{Baglio:2022wzu} and the N$^3$LO hard function \cite{Gehrmann:2010ue} contribution to the subtraction term are plotted, showing that these power corrections are sizable for $\tauc \gtrsim 10^{-5}$. 
 Note that the behaviour at very large rapidities is mainly due to the choice of normalization and will have a negligible effect in absolute terms.}
 \label{fig:Hgg_all_channels}
\end{figure*}
As we have seen in \Sec{ggH_analytic}, we have contributions from two channels. 
The numerical result for their sum is displayed in \fig{Hgg_all_channels}, together with the N$^3$LO inclusive K-factor from \cite{Baglio:2022wzu} and the N$^3$LO hard function contribution \cite{Gehrmann:2010ue} to the subtraction term for comparison.

For Drell-Yan we decided to show the results normalized by the leading order rapidity spectrum, since the higher order corrections are not large and therefore they do not change significantly the picture.
However, in the case of Higgs production in gluon fusion it is very well known that there are sizable higher order corrections, although they are pretty flat in rapidity \cite{Anastasiou:2004xq,Dulat:2017brz,Dulat:2017prg,Chen:2021isd}.
For this reason, normalizing N$^3$LO results by the LO rapidity spectrum would make all the N$^3$LO ingredients appear unnaturally large. 
While this equally affects both the calculated power corrections and the ingredients we are comparing them to, and thus wouldn't distort any numerical comparison between them, we preferred to avoid this.
Hence, we present the results in this section normalized by the product of the LO rapidity spectrum and the NNLO inclusive K-factor, which gives a good proxy for the size of the NNLO rapidity distribution we are computing corrections to.
Regarding the size of the LL power corrections, we see that, similarly to what we have seen in the previous section for Drell-Yan, they are quite sizable compared to the typical size of the N$^3$LO corrections, unless one considers $\tauc$ values significantly below $10^{-4}$, which is quite challenging numerically.
We also see that the dependence in rapidity is rather flat. 
Note that the behaviour at large rapidities is mainly caused by the vanishing of the normalization which is determined by the gluon PDF at large $x$.
This has an irrelevant effect in absolute terms since in that region there is very little cross section.

\begin{figure*}
 \centering
 \includegraphics[width=0.49\textwidth]{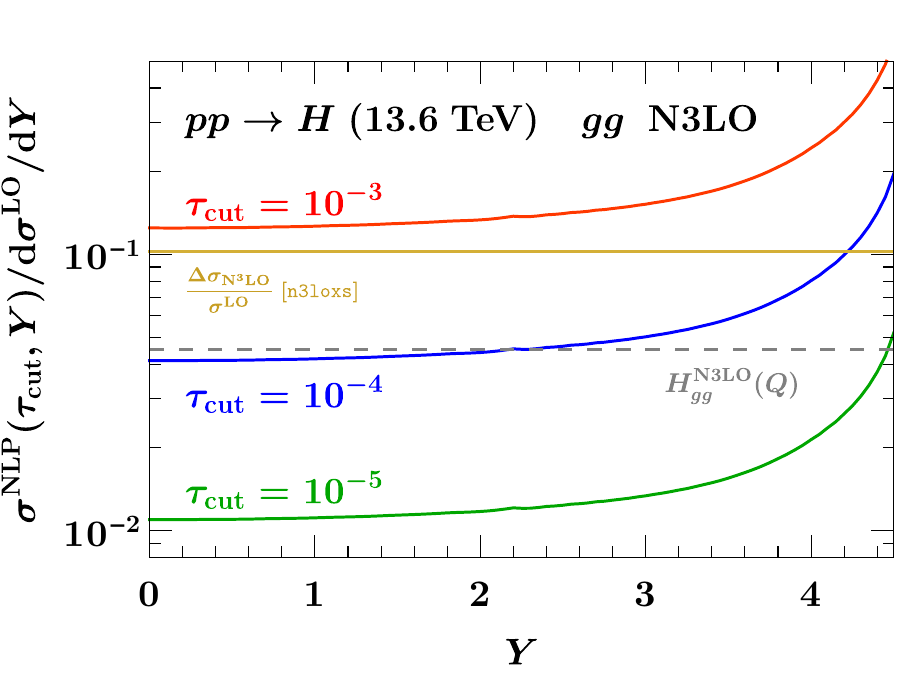}
 \hfill
 \includegraphics[width=0.49\textwidth]{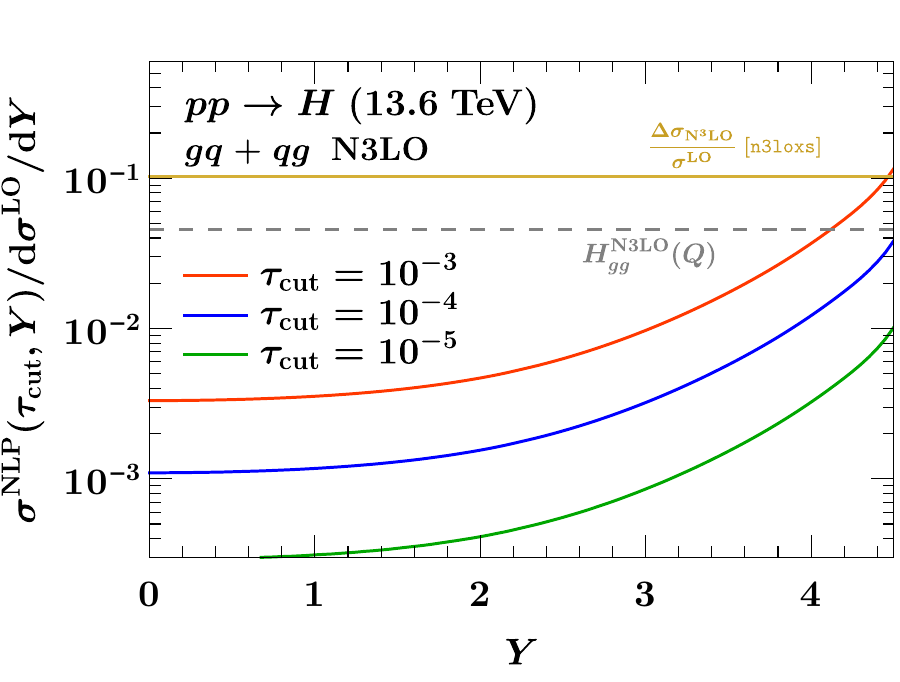}
 \caption{Same as \fig{DY_all_channels} but for different production channels. 
 We show the diagonal channel $q\bar{q}$ on the left and $qg$ on the right. 
 Both channels contribute at leading log at NLP. 
 It is evident that the power corrections to the diagonal channel $gg$ are the only one numerically relevant, since they constitute more than 95\% of the total contribution.
 Note that the behaviour at very large rapidities is mainly due to the choice of normalization and will have a negligible effect in absolute terms.}
 \label{fig:ggHchannel_decomposition}
\end{figure*}

For completeness, in \fig{ggHchannel_decomposition} we show the channel decomposition of our result.
We see that the contribution from the diagonal $gg$ channel is by far the most dominant one being more than an order of magnitude larger than the $gq$ channel.
This is in contrast with what we have seen in \Sec{DY_numerics} in the case of Drell-Yan where the off-diagonal channel was larger. 
Note that the $gq$ channel enters at the same logarithmic order as the diagonal channel and their relative size is most likely explained by the gluon PDF being large in the typical kinematic regions probed in Higgs production at the LHC.

\begin{figure*}
 \centering
 \includegraphics[width=0.8\textwidth]{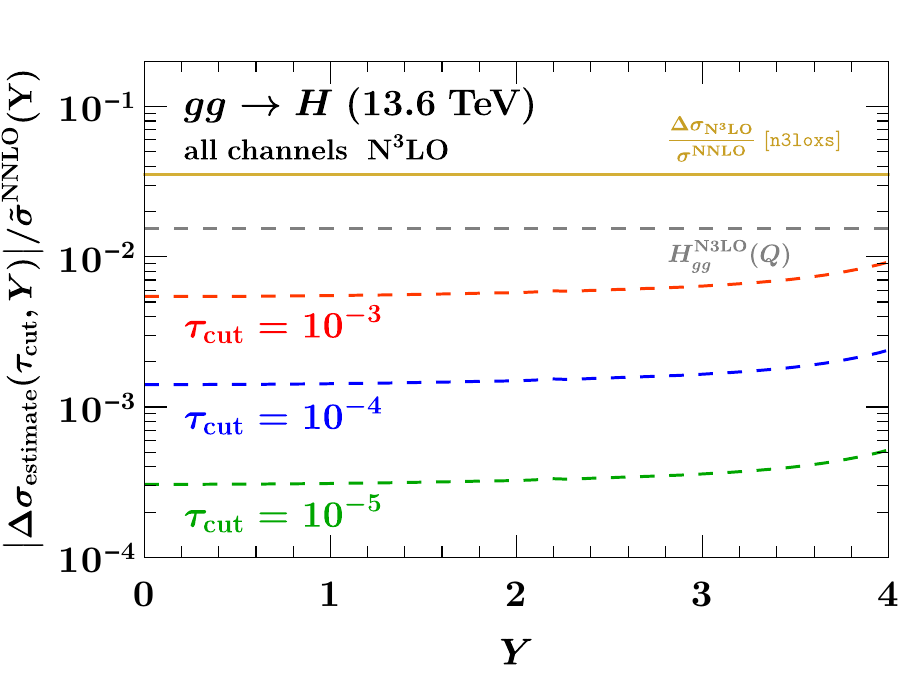}
 \caption{Estimate of the slicing error $\Delta \sigma(\tauc,Y)$ after the removal of the leading logarithmic terms analytically calculated in this work (i.e. after their inclusion in the subtraction term) for various values of $\tauc$. 
 Plotted for comparison are the inclusive N$^3$LO K-factor and the N$^3$LO hard function contribution to the subraction term.} 
 \label{fig:ggHresiduals}
\end{figure*}

Finally, in \fig{ggHresiduals} we repeat the exercise of estimating the size of the residual slicing uncertainties after including the results of \Sec{ggH_analytic} in the subtraction term, assuming a similar size between the coefficients of the logarithms at NLP.
We observe that the inclusion of these terms analytically alleviates the impact of slicing uncertainties and could allow for obtaining results with acceptable uncertainties with $\tauc$ not too far from $10^{-3}$.
If that turns out to be the case, it will be then interesting to see if this could roughly compete in terms of numerical efficiency with calculations purely based on P2B such as the one of \refcite{Chen:2021isd}.

\section{Conclusions}\label{sec:conc}
In this work, we calculated the leading logarithmic dynamical power corrections to the 0-jettiness distribution at next-to-next-to-next-to-leading order, differential in rapidity and invariant mass, for Drell-Yan and Higgs production in gluon fusion.
We have presented a method to disentangle dynamical power corrections, tied to QCD dynamics beyond leading power, from those arising from fiducial cuts using Projection to Born-improved subtractions.
From the calculation of these power corrections, we can estimate that if only the leading power terms in the 0-jettiness distribution are included in the subtraction term, it would be necessary to push the slicing parameter $\tauc$ to values below $10^{-5}$ to obtain results with slicing errors at few per-mille level. 
On the other hand, including these leading logarithmic terms in the subtraction can substantially improve the situation.
We estimate that removing this leading source of power corrections can yield a numerical improvement at N$^3$LO comparable to reducing the slicing parameter by $\cO(50)$ in the cases where there are no accidental large differences between the LL NLP coefficients and the ones multiplying the subleading logarithms.
This is the case at NLO and NNLO for Drell-Yan and gluon-fusion Higgs production at the LHC~\cite{Moult:2016fqy,Moult:2017jsg,Ebert:2018lzn}.
Therefore, we believe that with the inclusion of the leading logarithmic corrections and using the projection to Born improvement for the treatment of fiducial power corrections, it should be possible to perform fully differential N$^3$LO calculations with $\tauc \sim 10^{-4}-10^{-3}$ with slicing uncertainties under control.
This advancement opens the door to N$^3$LO predictions based on slicing with significantly enhanced efficiency, crucial for broadening the availability and utility of these predictions in precision phenomenology studies at current and future colliders.

\begin{acknowledgments}
We thank Alexander Huss for numerous discussions throughout this work, feedback on the manuscript, and for providing the numerical results from \refcite{Chen:2021vtu}.
We are grateful to Iain Stewart, HuaXing Zhu, and Ian Moult for discussions which initiated this project, to Alessandro Broggio and Simone Alioli for insightful discussions, and to Pier Monni for helpful comments on this work and feedback on the manuscript.
We thank the Erwin-Schrödinger International Institute for Mathematics and Physics at the University of Vienna for partial support during the Programme ``Quantum Field Theory at the Frontiers of the Strong Interactions". 
We thank the Galileo Galilei Institute for Theoretical Physics for the hospitality and the INFN for partial support during the completion of this work.
\end{acknowledgments}

\bibliography{subRGE}{}
\bibliographystyle{jhep}

\end{document}